\documentclass[submission,copyright,creativecommons,sharalike]{eptcs}
\providecommand{\event}{LFMTP 2018}
\usepackage{breakurl}
\usepackage[mathletters]{ucs}
\usepackage[utf8x]{inputenc}
\usepackage[frozencache]{minted}
\usepackage{fancyvrb}
\usepackage{amsmath}
\usepackage{bussproofs}
\usepackage{todonotes}
\usepackage{float}

\definecolor{bg}{rgb}{0.95,0.95,0.95}
\setminted[ocaml]{bgcolor=bg,fontsize=\footnotesize}
\floatstyle{boxed}
\restylefloat{figure}

\newcommand{\bnfeq}{\;\mathrel{::=}\;\;}
\newcommand{\bnfor}{\,\mathrel{\vert}\,}
\newcommand{\caml}[1]{\mintinline{ocaml}{#1}}

\VerbatimFootnotes

\title{Abstract Representation of Binders in OCaml\\
       using the Bindlib Library}

\author{Rodolphe Lepigre
\institute{Inria, LSV, CNRS, Université Paris-Saclay\\ Cachan, France}
\email{rodolphe.lepigre@inria.fr}
\and Christophe Raffalli
\institute{LAMA, CNRS, Université Savoie Mont Blanc\\ Chambéry, France}
\email{christophe.raffalli@univ-smb.fr}
}

\def\titlerunning{The Bindlib Library}
\def\authorrunning{R. Lepigre, C. Raffalli}

\begin{document}
\maketitle

\begin{abstract}
  The Bindlib library for OCaml provides a set of tools for the manipulation
  of data structures with variable binding. It is very well suited for the
  representation of abstract syntax trees, and has already been used for the
  implementation of half a dozen languages and proof assistants (including a
  new version of the logical framework Dedukti).
  Bindlib is optimised for fast substitution, and it supports variable
  renaming. Since the representation of binders is based on higher-order
  abstract syntax, variable capture cannot arise during substitution. As a
  consequence, variable names are not updated at substitution time. They can
  however be explicitly recomputed to avoid “visual capture” (i.e., distinct
  variables with the same apparent name) when a data structure is displayed.
\end{abstract}

\section{Introduction}

The implementation of programming languages and/or theorem provers plays an
important role in our community. It allows us to concretely illustrate the
mathematical models that we consider, and also helps us discovering new
intuitions through experimentations. In practice, implementing languages is
more difficult than one could expect. It requires the combination of specific
techniques, ranging from the parsing of source files to unification
algorithms, through the representation of data structures with variable
bindings. We here focus on the latter point and introduce the Bindlib
library~\cite{bindlib} for OCaml.

Variable binding is very common in computer science, as it is used for the
representation of computer programs and mathematical formulas as abstract
syntax trees. However, implementing the necessary primitives, including
variable renaming and capture-avoiding substitution, is cumbersome and
error-prone. Moreover, naive implementations generally result in poor
performances, especially when many substitutions must be performed. The
Bindlib library solves these problems thanks to an abstract representation of
binders with an efficient substitution operation. It is based on a form of
higher-order abstract syntax \cite{hoas},\footnote{The idea of higher-order
abstract syntax is to represent binders as functions in the host language. As
a consequence, the binding of a value of type \caml{'a} in a value of type
\caml{'b} is represented as a function of type \caml{'a -> 'b}.} which
eliminates the possibility of variable capture. Variable names are managed
using a distinct mechanism, and bound variables are not effectively renamed
when performing a substitution. As a consequence, displaying a data structure
in which substitutions have occurred may introduce “visual capture”, or
distinct variables with the same apparent name. To avoid this unfortunate, but
harmless situation, variable names must be explicitly updated.\footnote{Bound
variable names are changed in the minimal way, and they may still be
overloaded as in the \(λ\)-term \(λx.λx.x\) (which is \(α\)-equivalent to
\(λx.λy.y\)). It is possible to enforce a “stronger” form of renaming like
Barendregt's convention.} This is achieved using a glorified “identity
function”, which injects the data structure into a specific Bindlib type
constructor.

\subsection{Main working principles}

The Bindlib library provides type constructors \caml{'a var} and
\caml{('a,'b) binder}, for representing free variables (of type \caml{'a})
and binders (of a value of type \caml{'a} in an element of type \caml{'b})
respectively. A bound variable can be substituted using the \caml{subst}
function.
\begin{minted}{ocaml}
val subst : ('a,'b) binder -> 'a -> 'b
\end{minted}
However, it is not possible to bind a free variable directly, as there is
no built-in function of type \caml{'a var -> 'b -> ('a,'b) binder}. This is
not surprising because this function does not have any specific information
on the structure of the elements of type \caml{'b}.  To bind variables in a
data structure, it must first be injected in a specific type constructor
\caml{'a box}. Intuitively, an element of type \caml{'a box} corresponds to a
value of type \caml{'a} under construction, and its free variables can be
bound easily using \caml{bind_var}.
\begin{minted}{ocaml}
val bind_var : 'a var -> 'b box -> ('a,'b) binder box
\end{minted}
When an element of type \caml{'a box} has been fully constructed, and all the
desired variables have been bound, it can be “unboxed” to a value of type
\caml{'a} using the \caml{unbox} function.
\begin{minted}{ocaml}
val unbox : 'a box -> 'a
\end{minted}
In the process, its free variables are set to remain free. Indeed, binding
such variables would require lifting the value to the \caml{'a box} type
again, which is not very efficient as it requires a traversal of the data
structure.
Nonetheless, two representations of an element of type \caml{'a} must often
coexist and interact. The type \caml{'a} itself is used for pattern-matching
and for substituting binders, and the type \caml{'a box} is used whenever
variables need to be bound. The latter case arises not only at the
construction of a term, but also when one needs to work under binders (e.g.,
to compute the strong normal form of a \(λ\)-term).

The Bindlib library aims at being very flexible, and it does not impose any
restriction on the types of variables that can be bound, nor on the type of
elements in which they are bound. As a consequence, it is the responsibility
of the programmer to implement the lifting functions transforming values of
type \caml{'a} into values of type \caml{'a box} for every type \caml{'a}
that may contain bound variables.
This limited amount of boiler-plate code is generally straightforward, and
mainly amounts to providing \emph{smart constructors} for each constructor of
the abstract syntax tree. In other words, every constructor of the type
\caml{'a} must be lifted to the \caml{'a box} type using provided Bindlib
functions. In an earlier version of Bindlib, this process was partly automated
using an OCaml syntax extension. However, this approach proved to be rather
confusing for new users, and the benefits did not really outweigh the costs.

\subsection{Origins and applications of Bindlib}

The development of Bindlib \cite{bindlib} was initiated by the second author
in the nineties. An early version of the library was used to implement an
efficient normaliser for the \(λ\)-calculus \cite{normaliser}. It was then
made into a separate library, and used in the implementation of the first
version of the PML language \cite{pml}. Starting from version 4, Bindlib was
mostly rewritten by the first author, who proposed some simplifications an
made a significant documentation effort. Further simplifications and
improvements have been put into the version 5 of Bindlib \cite{bindlib},
which is being released at the time of writing. Bindlib is now
finally ready to be used by a larger community. The latest version
can be easily installed with the Opam package manager, using the following
command.
\begin{center}
\verb#opam update && opam install bindlib#
\end{center}

Bindlib has been and is used for the implementation of half a dozen
programming languages and/or proof assistants, as well as for a large
number of small prototypes, including an implementation of the pure type
systems (PTS), and an implementation of the combinatory reduction systems
(CRS). We give below a list of the most recent and relevant systems relying
on Bindlib.

\paragraph{The SubML language.}

The SubML language \cite{subml} implements a rich extension of System F with
Subtyping \cite{toplas}. It features polymorphic, existential, inductive and
coinductive types, which all require variable binding. Variable binding is
also used for the standard \(λ\)-abstraction constructor, as well as a
fixpoint for general recursion. The SubML language makes a particularly
extensive use of binders as the system relies on choice operators (similar
to Hilbert's Epsilon operator) in its syntax. Moreover, the language allows
quantification over sizes for inductive types, which requires binding a
syntactic representation of ordinals into types (as it is usually done for
sized types).

\paragraph{The PML\(_2\) proof system.}

The PML\(_2\) proof system \cite{pml2} implements the type system described in
the PhD thesis of the first author \cite{phd}. This project certainly contains
the most advanced uses of Bindlib so far, as it mixes many different types of
binders. In particular, the language admits a higher-order type system with
several base sorts (values, terms, stacks, propositions and ordinals), as
well as an arrow sort. In particular, Bindlib binders are mixed with GADTs in
order to implement multiple binders with non-homogeneous types (i.e., binding
values of different types at once into a term).

\paragraph{A new version of Dedukti.}

Bindlib was most recently used by the first author to propose a new
implementation of the logical framework Dedukti \cite{dedukti}, called
Lambdapi \cite{lambdapi}. Although the abstract syntax of the language
has a rather simple binding structure, with only \(λ\)-abstractions and
dependent product types, the system makes a singular use of Bindlib for
representing rewriting rules. A rewriting rule \(l \hookrightarrow r\) is formed of
a \emph{left-hand side} (LHS) \(l\), which corresponds to a pattern, and a
\emph{right-hand side} (RHS) \(r\), in which the free variables of
\(l\) are bound. For example, the term
“\(\mathtt{Plus}\;12\;(\mathtt{Succ}\;29)\)” matches the pattern of the
rewriting rule “\(\mathtt{Plus}\;n\;(\mathtt{Succ}\;m) \hookrightarrow
\mathtt{Succ}\;(\mathtt{Plus}\;n\;m)\)”, associating the values \(12\) and
\(29\) to variables \(n\) and \(m\), thus resulting in the term
“\(\mathtt{Succ}\;(\mathtt{Plus}\;12\;29)\)” after the
application of the rewriting rule. In the implementation, the RHS is
effectively represented as a binder ranging over the free variables of the
LHS. Applying a rule thus simply amounts to substituting this binder with
the values gathered for the variables during pattern-matching.

\subsection{Existing approaches to data structures with bound variables}

\paragraph{Naive approach.}

There exist several different approaches to the representation of data
structures with variable bindings. A first possibility is to use the naive
approach, which is very close to our pen and paper intuitions. In this
presentation, variables are represented with names, and they are bound by
simply referencing these names. For example, the pure \(λ\)-terms can be
encoded as follows.\footnote{Note that bound and free variables are
represented uniformly (using a single constructor), despite the fact that
these objects have very different status. This will not be the case in other
representations.}
\begin{minted}{ocaml}
(** Naive representation of pure λ-terms. *)
type term =
  | Var of string        (** Bound or free variable. *)
  | Abs of string * term (** Abstraction. *)
  | App of term * term   (** Application. *)
\end{minted}
Note that terms can be parsed and injected in the above data type in a
straightforward way. For example, the term \(λx.λy.x\;y\) is represented as
\caml{Abs("x", Abs("y", App(Var("x"), Var("y"))))}.  Although this
representation is convenient for the construction of terms, it is not very
well suited for implementation. Indeed, the capture-avoiding substitution
operation cannot be implemented efficiently, and it is also relatively hard to
get right. Overall, the main interest of the naive approach is purely
pedagogical, since it illustrates the usual notions of renaming and
capture-avoiding substitution at a low level.

\paragraph{De Bruijn indices}

The most widely used technique for implementing binders is certainly De Bruijn
indices \cite{debruijn}, in which bound variables are replaced with natural
numbers giving the “distance” between the variable and the linked
binder. The corresponding representation of pure \(λ\)-terms can be expressed
as follows.\footnote{The use of De Bruijn indices is sometimes referred to as
\emph{locally nameless}, referring to the fact that free variables are named.
This idea was already present in the work of De Bruijn, as remarked by
Charguéraud~\cite{nameless}.}
\begin{minted}{ocaml}
(** De Bruijn representation of pure λ-terms. *)
type term =
  | Var of string      (** Free variable.          *)
  | Ind of int         (** Index (bound variable). *)
  | Abs of term        (** Abstraction.            *)
  | App of term * term (** Application.            *)
\end{minted}
Using De Bruijn indices, the term \(λx.λy.x\;y\) is represented as
\caml{Abs(Abs(App(Ind(2),Ind(1))))}.\footnote{De Bruijn indices most often
start at \caml{0} for the immediate binder, but we here stick to \caml{1}
which was used in the original presentation \cite{debruijn}. Both approaches
are isomorphic, but using \caml{0} allows some optimisations.} Note that
\(α\)-equivalent terms have a unique representation using De Bruijn indices,
but that bound variable names are lost in the process (if they are not managed
using a specific mechanism).
Substitution using De Bruijn representation is well-defined, but hard to get
right in practice, especially if several kind of objects can be bound. Indeed,
the index to substitute (initially \caml{1}) increases when moving under
binders, and some index shifting is also necessary in the substituted term
when it contains indices that are bound outside the scope of the substitution.

\paragraph{Higher-order abstract syntax and its variations}

Another alternative is the use of higher-order abstract syntax (or
HOAS)~\cite{hoas}, in which a binder is represented using a function of the
host language. As it relies on the binders of the meta-language, the
correctness of this approach is immediate. A HOAS representation of pure
\(λ\)-terms is given below.
\begin{minted}{ocaml}
(** Higher-order abstract syntax representation of pure λ-terms. *)
type term =
  | Var of string         (** Free variable. *)
  | Abs of (term -> term) (** Abstraction.   *)
  | App of term * term    (** Application.   *)
\end{minted}
The term \(λx.λy.x\;y\) is here encoded as \caml{Abs(fun x -> Abs(fun y ->
App(x,y)))}.  As with De Bruijn indices, the name of bound variables are not
stored. As a consequence, they must be handled using a separate mechanism. A
variation of this technique is used in the internals of Bindlib, but it is not
visible to the user.

An important remark about the use of HOAS is that the domain of binders
appears negatively (i.e., to the left of an odd number of arrows). As a
consequence, the definition of the type \caml{term} above falls into
the limitations of languages such as Coq or Agda, where positivity is
enforced on inductive data types (this is required for their soundness). To
solve this issue, variations of HOAS such as \emph{parametric higher-order
abstract syntax}~\cite{phoas} or \emph{nested abstract syntax}~\cite{nested}
have been designed. Although they are more well-behaved than general HOAS (in
the sense that the corresponding types only contain terms of the represented
abstract syntax), such techniques are not very flexible.

\subsection{Related works and similar tools}

There are only very few available tools or libraries for the
representation of binders in the OCaml language. The only one that
seems to be a serious alternative is François Pottier's C\(α\)ml (or
alphaCaml)~\cite{alphacaml, alphacamlpaper}, which follows a completely
different approach. Indeed, C\(α\)ml is not an OCaml library, but a
meta-programming tool that generates OCaml code from a ``binding
specification''. Yet another difference with Bindlib is that C\(α\)ml
relies on De Bruijn indices, not a form of HOAS.

Another approach is that of the FreshML family of languages \cite{fresh3,
fresh2, fresh1}, which extend OCaml with specific binding facilities. There
seem to be some similarities between these tools and the Bindlib approach,
most notably in the idea of generating fresh names (or rather, variable) to
substitute binders. One advantage of FreshML over Bindlib is that the user
does not have to provide any code (with Bindlib, a lifting function must
be written by the user, as we will shortly see). However, this requires a
modification of the host language, while Bindlib is a simple library.


\section{Application to Church-style System F}

\begin{figure}
  Syntax of terms and types:
  \vspace{-\baselineskip}
  \begin{align*}
    t,u \bnfeq &x \bnfor λx:A.t \bnfor t\;u \bnfor ΛX.t \bnfor t\;A\\
    A,B \bnfeq &X \bnfor A ⇒ B \bnfor ∀X.A
  \end{align*}

  Operational semantics:
  \begin{prooftree}
    \AxiomC{}
    \UnaryInfC{\((λx:A.t)\;u \;\;\longrightarrow\;\; t[x:=u]\)}
    \DisplayProof\hfill
    \AxiomC{}
    \UnaryInfC{\((ΛX.t)\;A \;\;\longrightarrow\;\; t[X:=A]\)}
    \DisplayProof\hfill
    \AxiomC{\(t_1 \;\;\longrightarrow\;\; t_2\)}
    \RightLabel{\hspace{1mm}}
    \UnaryInfC{\(t_1\;u \;\;\longrightarrow\;\; t_2\;u\)}
  \end{prooftree}

  \smallskip

  \begin{prooftree}
    \AxiomC{\(u_1 \;\;\longrightarrow\;\; u_2\)}
    \UnaryInfC{\(t\;u_1 \;\;\longrightarrow\;\; t\;u_2\)}
    \DisplayProof\hspace{2cm}
    \AxiomC{\(t_1 \;\;\longrightarrow\;\; t_2\)}
    \UnaryInfC{\(λx:A.t_1 \;\;\longrightarrow\;\; λx:A.t_2\)}
    \DisplayProof\hspace{2cm}
    \AxiomC{\(t_1 \;\;\longrightarrow\;\; t_2\)}
    \RightLabel{\hspace{1mm}}
    \UnaryInfC{\(ΛX.t_1 \;\;\longrightarrow\;\; ΛX.t_2\)}
  \end{prooftree}

  Typing rules:
  \begin{prooftree}
    \AxiomC{}
    \UnaryInfC{\(Γ,\,x:A ⊢ x : A\)}
    \DisplayProof\hfill
    \AxiomC{\(Γ,\,x:A ⊢ t : B\)}
    \UnaryInfC{\(Γ ⊢ λx:A.t : A ⇒ B\)}
    \DisplayProof\hfill
    \AxiomC{\(Γ ⊢ t : A ⇒ B\)}
    \AxiomC{\(Γ ⊢ u : A\)}
    \RightLabel{\hspace{1mm}}
    \BinaryInfC{\(Γ ⊢ t\;u : B\)}
  \end{prooftree}

  \smallskip

  \begin{prooftree}
    \AxiomC{\(Γ ⊢ t : A\)}
    \AxiomC{\(X ∉ Γ\)}
    \BinaryInfC{\(Γ ⊢ ΛX.t : ∀X.A\)}
    \DisplayProof\hspace{4cm}
    \AxiomC{\(Γ ⊢ t : ∀X.A\)}
    \RightLabel{\hspace{1mm}}
    \UnaryInfC{\(Γ ⊢ t\;B : A[X := B]\)}
  \end{prooftree}

  \caption{Syntax, operational semantics and typing rules for Church-style
    System F.}\label{fig:fchurch}
\end{figure}
We will now consider the implementation of the Church-style (explicitly typed)
version of System F. This language is more interesting than the usual, pure
\(λ\)-calculus example because it requires the binding of types in types
(polymorphism), the binding of terms in terms (\(λ\)-abstractions), and the
binding of types in terms (type abstractions).
The syntax of the language, its operational semantics and its typing rules
are recalled in Figure~\ref{fig:fchurch}.

\subsection{Abstract syntax tree}

The abstract syntax tree of our language can be encoded as follow, using
the Bindlib data types. Free variables are represented using the
\caml{'a var} type, and binders using the \caml{('a,'b) binder} type.
\begin{minted}{ocaml}
(** Representation of a type. *)
type ty =
  | TyVar of ty var              (** Free type variable.   *)
  | TyArr of ty * ty             (** Arrow type.           *)
  | TyAll of (ty,ty) binder      (** Universal quantifier. *)

(** Representation of a term. *)
type te =
  | TeVar of te var              (** Free lambda-variable. *)
  | TeAbs of ty * (te,te) binder (** Lambda-abstraction.   *)
  | TeApp of te * te             (** Application.          *)
  | TeLam of (ty,te) binder      (** Type abstraction.     *)
  | TeSpe of te * ty             (** Type specialisation.  *)
\end{minted}
The definition of these types closely follows the BNF grammar given
at the top of Figure~\ref{fig:fchurch} page~\pageref{fig:fchurch}, and
similar definitions can be made for richer languages.

In the following sections, we will demonstrate the construction of terms
and types using this representation, and the implementation of various
functions. In particular, we will see how it is possible to work under
binders, using user-defined lifting functions. Note that such functions
are not needed when we do not want to work under binders, or when
binders only have to be substituted. For example, it is possible to define
the following head-normalisation function on terms.
\begin{minted}{ocaml}
(** Head-normalisation function. *)
let rec hnf : te -> te =
  function
  | TeApp(t,u) ->
      begin
        let v = hnf u in
        match hnf t with
        | TeAbs(_,b) -> hnf (subst b v)
        | h          -> TeApp(h,v)
      end
  | TeSpe(t,a) ->
      begin
        match hnf t with
        | TeLam(b) -> hnf (subst b a)
        | h        -> TeSpe(h,a)
      end
  | t          -> t
\end{minted}
On the other hand, we will only be able to implement a strong normalisation
function after the lifting operations (related to the construction of binders)
have been defined.

Of course, it is possible to implement variations of the head normalisation
function by changing the evaluation strategy, which is here right-to-left
call-by-value. It can also be implemented in the form of a stack machine
(e.g., a Krivine abstract machine).

\subsection{Smart constructors and lifting functions}

As discussed in the introduction, it is the responsibility of the programmer
to define the lifting functions for every type that may contain free
variables. Here, this is the case for both the \caml{te} and \caml{ty} types.
We will thus need to give two lifting functions which types will be \caml{te
-> te box} and \caml{ty -> ty box} respectively. To do so, it is a good
practice to first define a set of smart constructors, that will also be
useful for building terms and types.

Smart constructors are generally straightforward to define, as this only
requires lifting the corresponding constructors to boxed types. To this aim,
Bindlib provides functions such as the following.\footnote{Various similar
functions can be used to lift usual type constructor to their boxed forms. In
fact, they can all be implemented using two primitives, thanks to the fact
that the \caml{'a box} has an applicative functor structure (this will be
explained later).}
\begin{minted}{ocaml}
val box_apply  : ('a -> 'b) -> 'a box -> 'b box
val box_apply2 : ('a -> 'b -> 'c) -> 'a box -> 'b box -> 'c box
\end{minted}
Using these functions, we can define the following smart constructors for our
abstract syntax.
\begin{minted}{ocaml}
let _TyArr : ty box -> ty box -> ty box =
  box_apply2 (fun a b -> TyArr(a,b))

let _TyAll : (ty,ty) binder box -> ty box =
  box_apply (fun f -> TyAll(f))

let _TeAbs : ty box -> (te,te) binder box -> te box =
  box_apply2 (fun a f -> TeAbs(a,f))

let _TeApp : te box -> te box -> te box =
  box_apply2 (fun t u -> TeApp(t,u))

let _TeLam : (ty,te) binder box -> te box =
  box_apply (fun f -> TeLam(f))

let _TeSpe : te box -> ty box -> te box =
  box_apply2 (fun t a -> TeSpe(t,a))
\end{minted}
In the above, we did not provide any smart constructors for free variables.
They are handled using a specific function \caml{box_var},
which is used to make variables available
for binding.
\begin{minted}{ocaml}
val box_var : 'a var -> 'a box
\end{minted}
As every variable can potentially be bound in a boxed type, the variable
constructors of the abstract syntax do not make sense at this level. We
can however complete our set of smart constructors by using
synonyms of \caml{box_var}. This is obviously not necessary, but it makes the
use of smart constructors more uniform during the construction of terms.
\begin{minted}{ocaml}
let _TeVar : te var -> te box = box_var

let _TyVar : ty var -> ty box = box_var
\end{minted}

To be able to construct terms, we still need to introduce two Bindlib
functions. The first one is called \caml{new_var}, and allows the
creation of a new free variable. The second one is the \caml{bind_var}
function, which is used to effectively bind a variable in a boxed value. The
obtain boxed binder can then be fed to smart constructors like \caml{_TyAll}.
\begin{minted}{ocaml}
val new_var : ('a var -> 'a) -> string -> 'a var

val bind_var : 'a var -> 'b box -> ('a,'b) binder box
\end{minted}
Note that the \caml{new_var} function requires a \caml{string}, corresponding
to a preferred name for the bound variable, and a function for injecting a
\caml{'a var} into the \caml{'a} type.

Before defining the lifting functions for our abstract syntax representation
we will demonstrate the construction of types and terms. We will construct
the type \(∀X.∀Y.(X ⇒ Y) ⇒ X ⇒ Y\), and a corresponding application combinator
\(ΛX.ΛY.λf:X⇒Y.λa:X.f\;a\). They can be defined as follows.
\begin{minted}{ocaml}
(* Creation of variables. *)
let _X = new_var (fun x -> TyVar(x)) "X"
let _Y = new_var (fun x -> TyVar(x)) "Y"
let f  = new_var (fun x -> TeVar(x)) "f"
let a  = new_var (fun x -> TeVar(x)) "a"

(* Representation of the type X ⇒ Y. *)
let _X_arr_Y : ty box = _TyArr (_TyVar _X) (_TyVar _Y)

(* Representation of the type ∀X.∀Y.(X ⇒ Y) ⇒ X ⇒ Y. *)
let appl_ty : ty box =
  _TyAll (bind_var _X (_TyAll (bind_var _Y (_TyArr _X_arr_Y _X_arr_Y))))

(* Representation of the term ΛX.ΛY.λf:X⇒Y.λa:X.f a. *)
let appl_te : te box =
  _TeLam (bind_var _X (_TeLam (bind_var _Y (
    _TeAbs _X_arr_Y (bind_var f (_TeAbs (_TyVar _X) (bind_var a (
      _TeApp (_TeVar f) (_TeVar a)))))))))
\end{minted}
Note that the construction of these terms can then be completed by calling the
\caml{unbox} function, which type is \caml{'a box -> 'a}. It would then be
possible to pattern-match on these elements, as we did in the implementation
of the \caml{hnf} function.

The lifting functions can then be defined in a straightforward way using
the smart constructors. Note that the \caml{box_binder} function is used to
propagates the lifting operation under binders, provided a suitable lifting
function for their codomain. To do so, \caml{box_binder} simply substitutes
the given binder using a fresh variable, applies the lifting function, and
reconstructs the binder using the \caml{bind_var} function.
\begin{minted}{ocaml}
let rec lift_ty : ty -> ty box = fun a ->
  match a with
  | TyVar(x)   -> _TyVar x
  | TyArr(a,b) -> _TyArr (lift_ty a) (lift_ty b)
  | TyAll(f)   -> _TyAll (box_binder lift_ty f)

let rec lift_te : te -> te box = fun t ->
  match t with
  | TeVar(x)   -> _TeVar x
  | TeAbs(a,f) -> _TeAbs (lift_ty a) (box_binder lift_te f)
  | TeApp(t,u) -> _TeApp (lift_te t) (lift_te u)
  | TeLam(f)   -> _TeLam (box_binder lift_te f)
  | TeSpe(t,a) -> _TeSpe (lift_te t) (lift_ty a)
\end{minted}

\subsection{Basic functions: normalisation, printing and equality}

At the beginning of the current section, we were able to define a head
normalisation function \caml{hnf}, since it did not require working
under binders. We will now consider several examples of functions which
need to do so, starting with a strong normalisation.
\begin{minted}{ocaml}
let rec nf : te -> te = fun t ->
  match t with
  | TeVar(_)   -> t
  | TeAbs(a,f) ->
      let (x,t) = unbind f in
      TeAbs(a, unbox (bind_var x (lift_te (nf t))))
  | TeApp(t,u) ->
      let u = nf u in
      begin
        match nf t with
        | TeAbs(_,f) -> nf (subst f u)
        | t          -> TeApp(t,u)
      end
  | TeLam(f)   ->
      let (x,t) = unbind f in
      TeLam(unbox (bind_var x (lift_te (nf t))))
  | TeSpe(t,a) ->
      begin
        match nf t with
        | TeLam(f) -> nf (subst f a)
        | t        -> TeSpe(t,a)
      end
\end{minted}
Note that this function is similar to \caml{hnf} in most cases, except when
the term is a binder (i.e., when it is a \caml{TeAbs} or a \caml{TeLam}
constructor). To handle these, the corresponding binder is
substituted using a fresh variable using the \caml{unbind} function, which
returns a couple of the fresh variable and the term obtained after
substitution. It is then possible to call the normalisation function on this
term, and then reconstruct the binder. To do so, the term must be lifted using
our \caml{lift_te} function so that the variable \caml{x} is available for
binding. The binder is then reconstructed using the usual \caml{bind_var}
function, and it is then immediately unboxed.

Let us now consider the case of printing functions for our abstract
syntax. Although they need to work under binders, there is no boxing
involved since no binder must be reconstructed. We only provide the
printing function for the \caml{ty} type, as printing terms can
be achieved in a very similar way.\footnote{Note that we always add
parentheses in functions types to avoid ambiguities. It would of course
be possible to limit the number of such parentheses, but this is not
important here.}
\begin{minted}{ocaml}
let rec print_ty : out_channel -> ty -> unit = fun oc a ->
  match a with
  | TyVar(x)   -> output_string oc (name_of x)
  | TyArr(a,b) -> Printf.fprintf oc "(%a) ⇒ (%a)" print_ty a print_ty b
  | TyAll(f)   -> let (x,a) = unbind f in
                  Printf.fprintf oc "∀%s.%a" (name_of x) print_ty a
\end{minted}
As for \caml{nf}, the \caml{unbind} function is used to decompose binders
into pairs of a fresh variable and a term. However, the printing function
relies on \caml{name_of} to obtain the name of free variable. The name of
the variable \caml{x} returned by \caml{unbind} is built according to its
bound counterpart.

To conclude this section, we provide an equality function for the types
of our abstract syntax. As for printing, there is no boxing involved
because equality can be tested in a “destructive” way. The function
below relies on the \caml{eq_vars} for variable comparison (which
amounts to comparing their unique keys), and the function \caml{eq_binder}
is used for binders.
\begin{minted}{ocaml}
let rec eq_ty : ty -> ty -> bool = fun a b -> a == b ||
  match (a, b) with
  | (TyVar(x1)   , TyVar(x2)   ) -> eq_vars x1 x2
  | (TyArr(a1,b1), TyArr(a2,b2)) -> eq_ty a1 a2 && eq_ty b1 b2
  | (TyAll(f1)   , TyAll(f2)   ) -> eq_binder eq_ty f1 f2
  | (_           , _           ) -> false
\end{minted}
Note that \caml{eq_binder} simply substitutes the two given binders with the
same, fresh variable. The obtained bodies can then be compared using the given
function. In particular, the comparison is automatically performed modulo
\(α\)-equivalence in this way.

\subsection{Type-checking and type inference}

To complete our example, we will now consider the implementation of the
type system of Figure~\ref{fig:fchurch} page~\pageref{fig:fchurch}. We
will here represent a typing context as an association list, mapping free
term variables to types. Note that the corresponding lookup function needs
to use \caml{eq_vars} to compare the keys.
\begin{minted}{ocaml}
type context = (te var * ty) list

let find_ctxt : te var -> context -> ty option = fun x ctx ->
  try Some(snd (List.find (fun (y,_) -> eq_vars x y) ctx))
  with Not_found -> None
\end{minted}

The (mutually defined) type-inference and type-checking functions can then be
implemented as follows, using a standard bidirectional type-checking approach.
Note that Bindlib-specific functions only need to be used in the cases related
to binders. For instance, inferring the type of a \(λ\)-abstraction requires
a call to the \caml{unbind} function, and the returned variable is then used
to extend the context. In some cases, it is also necessary to establish a
binder. This is the case when inferring the type of a type abstraction, where
we start by decomposing the binder with \caml{unbind}, infer the type of the
body, and finally bind the variable returned by \caml{unbind} in the type to
produce a polymorphic type.
\begin{minted}{ocaml}
let rec infer : context -> te -> ty = fun ctx t ->
  match t with
  | TeVar(x)   ->
      begin
        match find_ctxt x ctx with
        | None    -> failwith "[infer] variable not in context..."
        | Some(a) -> a
      end
  | TeAbs(a,f) ->
      let (x,t) = unbind f in
      let b = infer ((x,a)::ctx) t in
      TyArr(a,b)
  | TeApp(t,u) ->
      begin
        match infer ctx t with
        | TyArr(a,b) -> check ctx u a; b
        | _          -> failwith "[infer] expected arrow type..."
      end
  | TeLam(f)   ->
      let (x,t) = unbind f in
      let a = infer ctx t in
      TyAll(unbox (bind_var x (lift_ty a)))
  | TeSpe(t,b) ->
      begin
        match infer ctx t with
        | TyAll(f) -> subst f b
        | _        -> failwith "[infer] expected quantifier..."
      end
\end{minted}

The type-checking function relies on similar techniques, and it uses the
previously defined \caml{eq_ty} function. In the case of a more expressive
language, like Martin-Löf dependent type theory, this equality function would
be replaced by a convertibility test that would involve evaluation (and hence
substitution). Examples of such situations can also be found in the
implementation of Lambdapi~\cite{lambdapi}, where conversion plays an even
greater role.
\begin{minted}{ocaml}
and check : context -> te -> ty -> unit = fun ctx t a ->
  match (t, a) with 
  | (TeVar(x)  , b         ) ->
      let a =
        match find_ctxt x ctx with
        | None    -> failwith "[check] variable not in context..."
        | Some(a) -> a
      in
      if not (eq_ty a b) then
        failwith "[check] type mismatch... (var)"
  | (TeAbs(c,f), TyArr(a,b)) ->
      if not (eq_ty c a) then
        failwith "[check] type mismatch... (abs)";
      let (x,t) = unbind f in
      check ((x,a)::ctx) t b
  | (TeApp(t,u), b         ) ->
      let a = infer ctx u in
      check ctx t (TyArr(a,b))
  | (TeLam(f1) , TyAll(f2) ) ->
      let (_,t,a) = unbind2 f1 f2 in
      check ctx t a
  | (TeSpe(t,b), a         ) ->
      begin
        match infer ctx t with
        | TyAll(f) ->
            let c = subst f b in
            if not (eq_ty c a) then
              failwith "[check] type mismatch... (spe)"
        | _        -> failwith "[infer] expected quantifier..."
      end
  | (_         , _         ) ->
      failwith "[check] not typable..."
\end{minted}

\section{Overview of the implementation of Bindlib}

We will now consider the core aspects of the implementation of the Bindlib
library, which can be downloaded at
\url{https://github.com/rlepigre/ocaml-bindlib}.
Note that the full library contains less than 1000 lines of generously
documented code.

\subsection{Main data structures}

At the core of Bindlib, the two most important data types are \caml{'a var}
(representing a free variable of type \caml{'a}) and \caml{'a box}
(representing an element of type \caml{'a} under construction). Their (mutual)
definitions are given and discussed below.
\begin{minted}{ocaml}
type 'a var =
  { var_key         : int          (** Unique identifier.                 *)
  ; var_prefix      : string       (** Prefix of the variable name.       *)
  ; var_suffix      : int          (** Suffix of the variable name.       *)
  ; var_mkfree      : 'a var -> 'a (** Free variable constructor in ['a]. *)
  ; mutable var_box : 'a box       (** Bindbox containing the variable.   *) }
\end{minted}

A free variable is uniquely identified by an integer, stored in the
\caml{var_key} field. It is not only used for referencing variables,
but also for comparing them. The name of a variable is split into a fixed
prefix, and an integer suffix which can be modified to avoid name clashes
when a variable is in the position of being bound. Note that a variable also
carries a function of type \caml{'a var -> 'a}, which is called for injecting
the (free) variable into the corresponding type. Conversely, a variable also
stores its boxed counterpart, so that it only needs to be computed
once.

\begin{minted}{ocaml}
and  'a box =
  | Box of 'a
  (** Element of type ['a] with no free variable. *)
  | Env of any_var list * int * 'a closure
  (** Element of type ['a] with free variables stored in an environment. *)
\end{minted}

A boxed element of type \caml{'a} is represented as \caml{Box(e)} in the
case where it does not contain any free variable. Otherwise, it is represented
as \caml{Env(vs,n,cl)}, which contains the list \caml{vs} of all the variables
that it effectively contains\footnote{In the current implementation, 
\caml{any_var} is defined as \caml{Obj.t var} and elements of \caml{'a var}
are coerced to \caml{any_var} using \caml{Obj.magic}. We could use an
existential type instead (as suggested by Bruno Barras), but this would
require an extra block of memory due to limitations on unboxing
(see \url{https://caml.inria.fr/mantis/view.php?id=7774}). Another advantage
of using an existential type would be that the \caml{var_box} field would not
need to be mutable anymore.} (sorted by key), an
integer \caml{n} giving the number of variables that have effectively been
bound, and the value represented as a closure \caml{cl}, which type is the
following.
\begin{minted}{ocaml}
type 'a closure = varpos -> Env.t -> 'a
\end{minted}
A closure expects two arguments. The former is a map, associating variable
keys to positions in the environment. The latter is the environment itself,
which is implemented as an array of heterogeneous values. The efficiency of
the substitution operation of Bindlib (\caml{subst} function) has to do with
the fact that closures are constructed in two steps. The first step only
involves the first argument of the closure, and consists in computing indices
in the environment. The second step only consists in accessing or manipulating
the environment, using the precomputed indices. In particular, the
\caml{varpos} map is only used once for each variable, even if the variable
appears many times.

\subsection{Variable manipulation and binding}

As mentioned in earlier sections, fresh variable are created using the
\caml{new_var} function, which has the type \caml{('a var -> 'a) -> string ->
'a var}. Its first argument is used as the value of the injection function
in the \caml{var_mkfree} field of the \caml{'a var} type, and the
\caml{string} name given as second argument is decomposed into a prefix and
an integer suffix for the \caml{var_prefix} and \caml{var_suffix} fields. The
variable key, stored in the \caml{var_key} field, is set using a new unique
identifier. Most interestingly, the \caml{var_box} field of the created
variable \caml{x} is first initialised with a dummy value, and then set as
follows.
\begin{minted}{ocaml}
let cl vp = Env.get (IMap.find var_key vp).index in
x.var_box <- Env([to_any x], 0, cl)
\end{minted}
The only variable in the box is \caml{x} itself, and \caml{0} variables were
bound. The closure computes the index corresponding to \caml{x} in the
environment, and then relies on \caml{Env.get} (of type \caml{int -> Env.t ->
'a}) to access the value in the environment in constant time.

We will now go into the construction of binders, which are created using the
\caml{bind_var} function, of type \caml{'a var -> 'b box -> ('a,'b)
binder box}. The type \caml{('a,'b) binder} itself is defined as follows, and
contains a name for the bound variable and a function of type
\caml{'a -> 'b} (in the spirit of higher-order abstract syntax), along with
several less important informations on the binder.
\begin{minted}{ocaml}
type ('a,'b) binder =
  { b_name   : string       (** Name of the bound variable.            *)
  ; b_bind   : bool         (** Indicates whether the variable occurs. *)
  ; b_rank   : int          (** Number of remaining free variables.    *)
  ; b_mkfree : 'a var -> 'a (** Injection of variables into domain.    *)
  ; b_value  : 'a -> 'b     (** Substitution function.                 *) }
\end{minted}
Note that the \caml{b_mkfree} field is initialised with the \caml{mk_free}
field of the variable being bound, so that the binder can be substituted with
a fresh variable when calling functions such as \caml{unbind}. 

The implementation of the \caml{bind_var} function cannot be explained in
full here, because it contains four different cases. In particular, the
computation of \caml{bind_var x t} depends on whether \caml{x} occurs in
\caml{t}, and whether if is the last free variable of \caml{t}. If \caml{x}
does not occur in \caml{t}, then it may be that \caml{t} is closed (i.e.,
of the form \caml{Box(v)}) or that \caml{x} is not in the list of its free
variables (i.e., \caml{t} is of the form \caml{Env(vs,n,cl)}, with \caml{x}
not in \caml{vs}). In both cases, the \caml{b_value} field of the binder is
set to be a constant function.
In the case where the variable occurs, the then term \caml{t} must be of
the form \caml{Env(vs,n,cl)} with \caml{x} appearing in \caml{vs}. In this
case, a position is reserved for the variable in the environment, and
associated to \caml{x.var_key} in the \caml{varpos} map.

\subsection{Boxing as an applicative functor}

The \caml{'a box} type constructor of Bindlib is an applicative functor which
unit is the \caml{box} function, and which application is the \caml{apply_box}
function. The former corresponds to an application of the \caml{Box}
constructor, and can be used to inject any value of type \caml{'a} in the
\caml{'a box} type, assuming that it does not contain any Bindlib variable.
The latter applies a boxed function to a boxed argument, obtaining a boxed
result. Note that these two primitives, along with \caml{box_var} and
\caml{bind_var} are the only four functions that are necessary for the
manipulation of boxed values, and the construction of binders. In particular,
\caml{box} and \caml{apply_box} can be combined to obtain convenient boxing
functions for usual data types. For instance, we can define the function
\caml{box_apply} and \caml{box_apply2} (that we used previously) as follows.
\begin{minted}{ocaml}
let box_apply : ('a -> 'b) -> 'a box -> 'b box =
  fun f a -> apply_box (box f) a

let box_apply2 : ('a -> 'b -> 'c) -> 'a box -> 'b box -> 'c box =
  fun f ta tb -> apply_box (box_apply f ta) tb
\end{minted}
We can also define boxing functions for the \caml{'a option} type or the
\caml{'a list} type as follows.
\begin{minted}{ocaml}
let box_opt : 'a box option -> 'a option box =
  function
  | None    -> box None
  | Some(e) -> box_apply (fun e -> Some(e)) e

let box_list : 'a box list -> 'a list box =
  fun l -> List.fold_right (box_apply2 (fun x l -> x::l)) l (box [])
\end{minted}

In the implementation, some of the provided boxing functions are not defined
from \caml{box} and \caml{box_apply} for performance reasons. This is the case
for the \caml{box_list} function, which is implemented using a more general
and optimised (imperative) functor mechanism, which only requires a \caml{map}
function.

\subsection{On the use of magic}

The implementation of Bindlib relies on the \caml{Obj} module\footnote{This
module gives access to the low-level representation of data structures, and
allows us to bypass type-checking.} of OCaml to store values of different
types in a single array, intuitively corresponding to the environments of a
closure. However, the implementation of Bindlib satisfies the invariant that
the value at every given index is always read or written at a fixed type. As
mentioned earlier, the \caml{Obj.magic} function is also used to transform
variables of type \caml{'a var} into value of type \caml{any_var} (defined
as \caml{Obj.t var}), which are only used in a type-safe way.

Although the lists of variables with heterogeneous types used in the \caml{'a
box} type could be encoded with an existential type, this is not the case for
the environments. Ensuring (more) type safety in the handling of environments
would require the user to provide a runtime representation of all the type
of variables that can be bound. We chose not to pursue this directions as
this would make the use of Bindlib cumbersome, and the only benefit would be
to have an exception raised instead of a segmentation fault, which could only
happen in case of a Bindlib bug.

\section{Conclusion and future work}

The Bindlib library has been around for more than twenty years, but it was
only used in a very restricted community centered around the second author.
The main reason for that was that the library was not very well documented,
and hence not easily accessible to potential users. This problem has now
been fixed, and the library is now extensively documented.

In a recent work, Bruno Barras has initiated the implementation of a Coq
version of Bindlib. Although it is not publicly available yet, this work opens
the way to the validation of the Bindlib model with a formal specification and
formal proofs. Due to the specificities of Bindlib, this requires an axiomatic
presentation, notably for the representation and the manipulation of
environments.

\nocite{*}
\bibliographystyle{eptcs}
\bibliography{biblio}
\end{document}